\documentclass{INTERSPEECH2023}

\usepackage{verbatim}
\usepackage{multirow}
\usepackage{makecell}
\title{Adapting OpenAI's Whisper for Speech Recognition on Code-Switch
Mandarin-English SEAME and ASRU2019 Datasets
}
\name{Yuhang Yang$^1$, Yizhou Peng$^2$, Xionghu Zhong$^1$, Hao Huang$^3$, Eng Siong Chng$^4$}
\address{
  $^1$ Hunan University, China\\
  $^2$ National University of Singapore, Singapore \\
  $^3$ School of Computer Science and Engineering, Xinjiang University, China \\
  $^4$ Nanyang Technological University, Singapore}
\email{}
\interspeechcameraready
\begin{document}

\maketitle
 
\begin{abstract}
This paper details the experimental results of adapting the OpenAI's Whisper model for Code-Switch Mandarin-English Speech Recognition (ASR) on the
SEAME and ASRU2019 corpora. We conducted 2 experiments:
a) using adaptation data from 1 to 100/200 hours to demonstrate effectiveness of adaptation, b) examining different language ID setup on Whisper prompt. 

The Mixed Error Rate results show that the amount of adaptation data may be as low as $1\sim10$ hours to achieve saturation in performance gain (SEAME) while the ASRU task continued to show performance
with more adaptation data ($>$100 hours). 
For the language prompt, the results show that although various prompting strategies initially produce different outcomes, adapting the Whisper model with code-switch data uniformly improves its performance.

These results may be relevant also to the community when applying Whisper for related tasks of adapting to new target domains.


\end{abstract}

\section{Introduction}
\label{sec:intro}
Code-switching is a pervasive linguistic phenomenon in multilingual communities, where speakers alternate between two or more languages within a single speech or conversation. The prevalence of code-switching poses a unique challenge for Automatic Speech Recognition (ASR) systems~\cite{Guodong-ASRU2023,CS-Mine}, as it requires the model to have a nuanced understanding of multiple languages simultaneously. Unfortunately, current ASR models often underperform in recognizing code-switching speech due to the scarcity of labeled training data. 

In the past a few years, the research community has introduced several approaches to tackle the challenges posed by code-switching ASR systems. These approaches can be broadly categorized into three technical aspects: speech, text, and modeling methods. From the speech perspective, strategies have been developed to implement monolingual speech into code-switching ASR systems~\cite{CS-Speech-1}. Additionally, some researchers propose augmenting pronunciation models to accommodate accents, mispronunciations, and pronunciation variations, thereby addressing the issue of data sparsity~\cite{CS-Speech-2}. On the text front, multiple techniques have been explored, ranging from augmenting code-switching text from monolingual corpora to build language models (LM)~\cite{CS-Text-1,CS-Text-2,CS-Text-3,CS-Text-4,CS-Text-5,CS-Text-6}, to employing methods like speech T5~\cite{CS-Text-7,CS-Text-8}, multilingual word-embedding~\cite{CS-Word-1,CS-Word-2,CS-Word-3,CS-Word-4,CS-Word-5}, Internal LM estimation~\cite{CS-LM-1,CS-LM-2,CS-LM-3}, and LM rescoring~\cite{CS-LM-4}. Lastly, from the modeling standpoint, various frameworks have been suggested, such as the Mixture of Experts (MoE) which uses separate encoders and decoders for different languages~\cite{CS-MOE-1,CS-MOE-2}, frame-level Language Identification or Diarization as an auxiliary task~\cite{CS-LID-1,CS-LID-2}, and the incorporation of self-supervised models as frontend models for ASR~\cite{CS-SSL-1}.


When superlarge parameter models show their emergent ability to understand and generate language when given a suitable prompt~\cite{Prompt-LFM}, researchers are turning to large-scale foundational models trained on extensive multilingual datasets, e.g., Whisper~\cite{Whisper}, USM~\cite{USm} and MMS~\cite{MMS}. These models aim to encapsulate a wide range of linguistic rules and contexts, offering more robust performance across different languages and dialects. Using large-scale training data and advanced modeling techniques, these foundational models have the potential to revolutionize the field of ASR, making it more inclusive and accurate for multilingual and code-switching populations. 

In this paper, we concentrate on the application of varied language labels as prompts during both the training and decoding phases of the Whisper model. Our investigation centers on the efficacy of these language label prompts in the finetuning process of the model. Additionally, we propose a prompting approach that considers code-switching as a distinct language. This method derives language embeddings through a weighted combination of the respective language embeddings, such as Mandarin and English, attempts to enhance model performance under code-switching scenario.

The paper is organized as follows. Section~\ref{sec:related} is to review recently proposed adaptation methods based on the Whisper model to Code-Switching or specific languages. Section~\ref{sec:proposed} presents the methods that are used in our experiments. Section~\ref{sec:exp} briefly summarizes the datasets used and the overall experimental setup. Section~\ref{sec:Results} shows our experimental results and section~\ref{sec:Abl} shows our ablation study on the size of the training data. After that, we draw conclusions in Section~\ref{sec:conclusion}.

\begin{table*}[htp]
 \centering
\caption{Prompt used in PromptingWhisper for CS-ASR and our proposed Language-Fusion Prompt for both finetuning and decoding. \textnormal{\textlangle \textbar sot\textbar \textrangle}~stands for  \textnormal{\textlangle \textbar startoftranscript\textbar \textrangle}~token and \textnormal{\textlangle \textbar asr\textbar \textrangle}~means the  \textnormal{\textlangle \textbar transcribe\textbar \textrangle} token in Whisper Tokenizer.}
 \label{tab:prompt-egs}
\begin{tabular}{cccc}
 \toprule
 Languages & Default & PromptingWhisper & Language-Fusion \\
   \midrule
   Zh+En & \textlangle \textbar sot\textbar \textrangle \textbf{\textlangle \textbar zh\textbar \textrangle}~\textit{or} \textbf{\textlangle \textbar en\textbar \textrangle}\textlangle \textbar asr\textbar \textrangle & \textlangle \textbar sot\textbar \textrangle \textbf{\textlangle \textbar zh\textbar \textrangle \textlangle \textbar en\textbar \textrangle}~\textit{or}  \textbf{\textlangle \textbar en\textbar \textrangle \textlangle \textbar zh\textbar \textrangle}\textlangle \textbar asr\textbar \textrangle & \textlangle \textbar sot\textbar \textrangle \textbf{\textlangle \textbar en-zh\textbar \textrangle}\textlangle \textbar asr\textbar \textrangle\\ 

    \bottomrule
\end{tabular}
\end{table*}

\section{Whisper Applications}
\label{sec:related}

The Whisper model~\cite{Whisper} is an advanced speech recognition system created by OpenAI that can transcribe audio into text with high accuracy. It is trained on a wide-ranging dataset, enabling it to handle multiple languages and dialects effectively, capable of speech translation and language identification except for speech recognition.

Whisper stands out for its robust performance in various acoustic settings and its contextual understanding of improved transcription, which makes it distinct for various speech and text research tasks that are benefits from Whisper encoder and decoder separately or simultaneously. Specifically, following the same training pipeline and finetuning the entire Whisper model, performance improvement is obtained for several low-resource languages speech recognition~\cite{Vistaar,N-Shot}. TCPGen also shows it's effectiveness of contextual biasing for Whisper model~\cite{Contextual-Biasing}. 
With the utilization of the Whisper encoder, deep-fake detection can benefit from input features that are composed of the embeddings extracted from the last layer of Whisper encoder and traditional MFCC features~\cite{deepfake}. With a similar strategy applied, these embeddings can also improve the performance of infant cry classification models compared to the MFCC features~\cite{inf-cry-classification}. Also, researchers indicate that the embeddings from the Whisper encoder are not only noise robust to ASR task, but also contains information that can help classfity the noise types, e.g., audio event tagging~\cite{Whisper-AT}. 
When it comes to the Whisper decoder, researchers usually try to prompt the decoder for various applications. By simply replacing the speech recognition transcription label with several Spoken Language Understanding (SLU) labels while keeping the same decoding prompt as ASR task, Whisper can perform SLU through transfer learning and multitask learning~\cite{WhiSLU}. Also, when customized prompts are fed into the Whisper decoder, improved performance is discovered for several zero-shot tasks such as Code-Switching ASR and Speech Translation that were previously underperformed with the default Whisper prompts~\cite{PromptingWhisper}.

\begin{figure}[htb]
    \centering
    \includegraphics[width=0.95\linewidth]{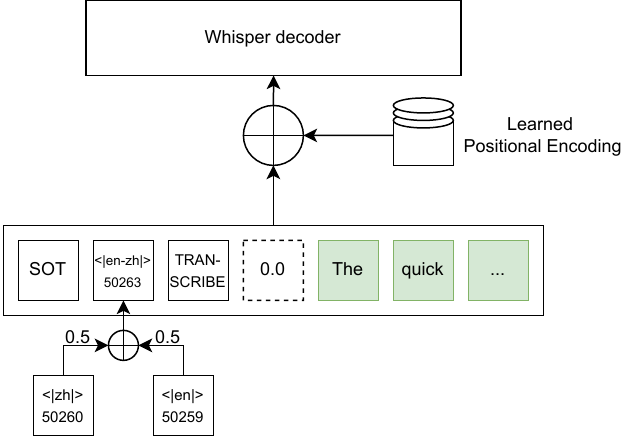}
    \caption{The proposed Language Fusion method}
    \label{fig:Language-Fusion}
    \vspace{-0.9em}
\end{figure}

\section{Methods}
\label{sec:proposed}

\subsection{PromptingWhisper for CS-ASR}
\label{ssec:promptingwhisper}
\textbf{PromptingWhisper}~\cite{PromptingWhisper} is an innovative approach in the field of speech recognition, focusing on the adaptation of Whisper model to new and untrained tasks through the technique of prompt engineering. This method involves the strategic use of prompts - specific instructions or inputs - to guide the Whisper model in processing and responding to tasks beyond its original training. PromptingWhisper primarily explores three tasks: audio-visual speech recognition (AVSR), where the model transcribes speech from videos with related visual content; code-switched speech recognition (CS-ASR), which involves recognizing speech that alternates between different languages; and speech translation (ST) for language pairs that the model has not previously encountered. 

For CS-ASR, PromptingWhisper suggests combining two Language Prompts instead of using only one of the two Languages, e.g., Table~\ref{tab:prompt-egs} shows all the prompts for Mandarin-English CS-ASR. Precisely, following the default prompting rule of Whisper~\cite{Whisper}, the resulting prompt for speech recognition would be either \textlangle \textbar sot\textbar \textrangle \textbf{\textlangle \textbar zh\textbar\textrangle}\textlangle \textbar asr\textbar \textrangle~or \textlangle \textbar sot\textbar \textrangle \textbf{\textlangle \textbar en\textbar \textrangle}\textlangle \textbar asr\textbar \textrangle~where only one language should be specified, however, PromptingWhisper suggest that given prompt like \textlangle \textbar sot\textbar \textrangle \textbf{\textlangle \textbar zh\textbar \textrangle \textlangle \textbar en\textbar \textrangle}\textlangle \textbar asr\textbar \textrangle~where sequentially inserting two languages would introduce 19\% relative performance improvement in average for CS-ASR task among several CS corpora such as SEAME~\cite{SEAME} and ASCEND~\cite{ASCEND}. Likewise, we think \textlangle \textbar sot\textbar \textrangle \textbf{\textlangle \textbar en\textbar \textrangle \textlangle \textbar zh\textbar \textrangle}\textlangle \textbar asr\textbar \textrangle which only reversing the order of two languages should also have similar outcome. Therefore, we apply both \textbf{\textlangle \textbar en\textbar \textrangle \textlangle \textbar zh\textbar \textrangle}~and \textbf{\textlangle \textbar zh\textbar \textrangle \textlangle \textbar en\textbar \textrangle} language prompts in the following experiments, treating them as PromptingWhisper's suggested prompts.

\subsection{Language-Fusion Prompt}
\label{ssec:lf} 
Inspired by the \textbf{PromptingWhisper}, we introduce a new Language Prompt that fuses the pretrained English and Chinese Language Embeddings, which is named as \textbf{Language-Fusion Prompt}, to investigate if this could be beneficial to English-Mandarin CS-ASR task. Typically, as shown in Figure~\ref{fig:Language-Fusion}, the new embedding is obtained by weighting the pretrained embeddings corresponding to English and Chinese Language Prompt tokens that are \textbf{\textlangle \textbar en\textbar \textrangle}~(\textit{Token id in Whisper Tokenizer is 50259}) and \textbf{\textlangle\textbar zh\textbar \textrangle}~(\textit{Token id in Whisper Tokenizer is 50260}) with the same weighting factor which is set to 0.5. In order to keep the same output dimension as of original Whisper Decoder, the \textbf{\textlangle\textbar ru\textbar\textrangle}~(\textit{Stands for Russian Language. Token id in Whisper Tokenizer is 50263}) is then replaced with the resulting new Language Prompt which is named as \textbf{\textlangle\textbar en-zh\textbar\textrangle}. The reason we select Russian language prompt as the replacement is from the substantial acoustic and linguistic differences between Russian and English/Chinese. The decoding prompt is finalized as \textlangle \textbar sot\textbar \textrangle \textbf{\textlangle \textbar en-zh\textbar \textrangle}\textlangle \textbar asr\textbar \textrangle shown in Table~\ref{tab:prompt-egs}.

\begin{table}[htp]
 \centering
\caption{Overall Speech Data distribution for both ASR model training and testing. \textnormal{Dev\textsubscript{Man}} and \textnormal{Dev\textsubscript{Sge}} are two officially defined testsets for SEAME corpus where the first is dominated with Mandarin and vice versa, the other is dominated with English. All ASRU datasets are dominated with Mandarin.}
 \label{tab:speech-data}
\begin{tabular}{cccc}
 \toprule
  Corpus & Subset & Duration(Hrs)  
  \\
   \midrule
   \multirow{3}*{SEAME} & Train & 93.6 \\
   & Dev\textsubscript{Man} & 7.5 \\
   & Dev\textsubscript{Sge} & 3.9 \\
\midrule
    \multirow{4}*{ASRU} & Train & 193.0 \\
    & Dev1 & 20.4 \\
    & Dev2 & 21.3 \\
    & Test & 20.6 \\

    \bottomrule
\end{tabular}
\end{table}
\section{Experiments}
\label{sec:exp}

\subsection{Data}

\label{subsec:data}

We select two Mandarin-English code-switching ASR data sets to verify the effectiveness of all the methods mentioned in Section~\ref{sec:proposed}. One is SEAME~\cite{SEAME}, a conversational Mandarin-English corpus from SouthEast Asia, i.e., Malaysia and Singapore. Another is a Mandarin-English CS data set from Chinese Mainland, released by \texttt{Datatang} for a Mandarin-English CS ASR challenge in ASRU2019~\cite{ASRU2019}. For brevity, we name it as ASRU in what follows.
Though both data sets are Mandarin-English CS, they are hugely different. Firstly, they are from different areas which means CS influenced with different cultural background.
More importantly, SEAME data is conversational speech, while ASRU is reading speech, and hence much simpler. Table~\ref{tab:speech-data} reports overall speech data distributions in details.

\subsection{Model}
\label{subsec:model}

All of our experiments are performed with the Whisper-small multilingual model due to the limitation of computing resources. The encoder is configured with 12 layers, and the decoder consists of 12 layers with 8-head attention. The input feature is 80-dim Mel frequency bins which is computed on 25-ms windows with a stride of 10 ms. 
All of our models are trained on one A40 GPU with 48GB VRAM. The original batch size is set to 6 and the gradient accumulation is 12, which results in batch size of 72 in total. 
We use AdamW optimizer with the peak learning rate of $1e^{-5}$, and the warmup lasts for 200 steps. The max updating steps is set to 30k. 
Also, mixed-precision training strategy~\cite{Mix-Precision} is applied in our experiments. 
When decoding, we use average model from last 5 epochs, and decoding beamsize is set to 1.

\section{Results}
\label{sec:Results}
The experimental results include mainly two parts. First, we show the results before finetuning the Whisper model, which is called the Zero-shot Prompt. Then, by finetuning the Whisper model following different Prompt styles, we show the huge improvements for all Prompt styles among the two Code-Switching datasets. 
\begin{table}[htp]
 \centering
\caption{MERs(\%) with different Language Prompt for Zero-Shot Code-Switching ASR. Conformer follows the configuration from recipe~\cite{ESPNet2-SEAME} in ESPnet2 toolkit. \textbf{Official} type represents to original Whisper Language Prompt style, where \textnormal{\textlangle \textbar en\textbar \textrangle} and \textnormal{\textlangle \textbar zh\textbar \textrangle} stand for specifying English and Mandarin for the entire testset respectively. \textit{Auto} denotes that we don't manually state the Language Prompt when perform decoding, means that Whisper would automatically recognize the Language Label for each sentence. \textbf{Custom} type includes two combined Language Prompts following PromptingWhisper where \textnormal{\textlangle \textbar en\textbar \textrangle \textlangle \textbar zh\textbar \textrangle} stands for English first and Mandarin second in the combined Language Prompt and vice versa. The proposed weighted-sum method is shown as \textnormal{\textlangle \textbar en-zh\textbar \textrangle}. }
 \label{tab:zero-shot-results}
 \begin{tabular}{ccp{18pt}p{18pt}|p{10pt}p{10pt}p{10pt}}
 \toprule
 \multirow{2}{*}{Type} & \multirow{2}{*}{L-Prompt} & \multicolumn{2}{c|}{SEAME} & \multicolumn{3}{c}{ASRU}  \\
  & &  Dev\textsubscript{Man} &  Dev\textsubscript{Sge} & \centering Dev1 &  Dev2 & Test \\
  
 \midrule
  Conformer & N/A  & \textbf{16.6} & \textbf{23.3} & \textbf{8.6} & \textbf{14.0} & \textbf{13.2} \\
  \midrule
 \multirow{3}{*}{Official} & \textlangle \textbar en\textbar \textrangle & 101.9 & 83.6 & 96.0 & 98.9 & 105.3 \\   
    & \textlangle \textbar zh\textbar \textrangle & 80.8 & 157.5 & 27.0 & 25.3 & 25.0 \\
  & \textit{Auto} & 67.8 &  84.9 & 31.1 &  29.9 &  29.4 \\
  \midrule
  \multirow{3}{*}{Custom} & \textlangle\textbar en\textbar\textrangle \textlangle\textbar zh\textbar\textrangle   & 84.0 & 81.3 & 98.4 & 101.1 & 99.4 \\
  & \textlangle \textbar zh\textbar \textrangle \textlangle \textbar en\textbar \textrangle  & 98.8 & 81.2 & 33.2 & 32.2 & 32.3 \\
  & \textlangle \textbar en-zh\textbar \textrangle & 74.0 & 101.7 & 33.8 & 31.9 & 31.6 \\
\bottomrule
\end{tabular}
\end{table}
\subsection{Zero-Shot Prompts on Whisper-Small}
We follow the instructions of PromptingWhisper and apply the suggested Language Prompts as well as our proposed method to Whisper-small,and the results are shown in Table~\ref{tab:zero-shot-results}. 

The results show that for all testsets, the Whisper-small model with various Language-Prompts underperforms Conformer models that are trained with corresponding training data. However, different Language-Prompts do significantly affect the performance of Code-Switching speech recognition. 
Specifically, for SEAME dataset, \textbf{\textlangle \textbar en\textbar \textrangle}~prompt gives better result for Dev\textsubscript{Sge} while \textbf{\textlangle \textbar zh\textbar \textrangle}~shows better performance on Dev\textsubscript{Man}. The \textit{Auto} prompt shows better performance compared with those with specific language prompt for Dev\textsubscript{Man} and similar result for Dev\textsubscript{Sge}. When given customized prompt, \textbf{\textlangle\textbar en\textbar\textrangle \textlangle\textbar zh\textbar\textrangle}~and \textbf{\textlangle\textbar zh\textbar\textrangle \textlangle\textbar en\textbar\textrangle}~prompts show best results on Dev\textsubscript{Sge}, while Dev\textsubscript{Man} underperforms the official prompts. Our proposed Language-Fusion prompt  \textbf{\textlangle \textbar en-zh\textbar \textrangle}~also does not perform well in zero-shot scenario.
When it comes with ASRU dataset, \textbf{\textlangle \textbar zh\textbar \textrangle}~and \textit{Auto} prompts show significant improvement in performance compared with \textbf{\textlangle \textbar en\textbar \textrangle ~and \textlangle\textbar en\textbar\textrangle \textlangle\textbar zh\textbar\textrangle}~prompts which shows consistency to data composition of ASRU. 



\subsection{Finetuning Whisper Model}
We finetune the Whisper-Small model given different language prompts and then specify exactly the same language prompt when performing decoding. Table~\ref{tab:finetune-results} shows the MER results of all Whisper models we finetuned on SEAME and ASRU datasets.

The results show that regardless of which Language Prompt we use for finetuning and decoding, all the testsets obtain significant performance improvements and outperform the Conformer models that are trained with corresponding training set.
\begin{table}[htp]
 \centering
\caption{MERs(\%) with different Language Prompt for Whisper-small finetuned model. \textnormal{\textlangle \textbar ru\textbar \textrangle} stands for Russian Language Prompt.}
 \label{tab:finetune-results}
 \begin{tabular}{ccp{18pt}p{18pt}|p{10pt}p{10pt}p{10pt}}
 \toprule
 \multirow{2}{*}{Type} & \multirow{2}{*}{L-Prompt} & \multicolumn{2}{c|}{SEAME} & \multicolumn{3}{c}{ASRU}  \\
  & &  Dev\textsubscript{Man} &  Dev\textsubscript{Sge} & \centering Dev1 &  Dev2 & Test \\
  
 \midrule
  Conformer & N/A  & \textbf{16.6} & {23.3} & {8.6} & {14.0} & {13.2} \\
  \midrule
 \multirow{3}{*}{Official} & \textlangle \textbar en\textbar \textrangle & \textbf{14.3} & \textbf{20.4} & \textbf{6.3} & 10.9 & 10.3 \\   
    & \textlangle \textbar zh\textbar \textrangle & 14.8 & 20.6 & \textbf{6.3} & 10.8 & \textbf{10.1} \\
  & \textbf{\textlangle \textbar ru\textbar \textrangle} & 15.5 &  21.5 & 6.3 &  10.8 &  10.3 \\
  \midrule
  \multirow{3}{*}{Custom} & \textlangle\textbar en\textbar\textrangle \textlangle\textbar zh\textbar\textrangle   & 15.1 & 21.1 & \textbf{6.3} & \textbf{10.6} & \textbf{10.1} \\
  & \textlangle \textbar zh\textbar \textrangle \textlangle \textbar en\textbar \textrangle  & 15.0 & 21.0 & 6.5 & 11.0 & 10.5 \\
  & \textlangle \textbar en-zh\textbar \textrangle & 15.1 & 20.9 & \textbf{6.3} & 10.8 & \textbf{10.1} \\
\bottomrule
\end{tabular}
\end{table}

\begin{figure*}[htb]
    \centering
    \includegraphics[width=0.95\linewidth]{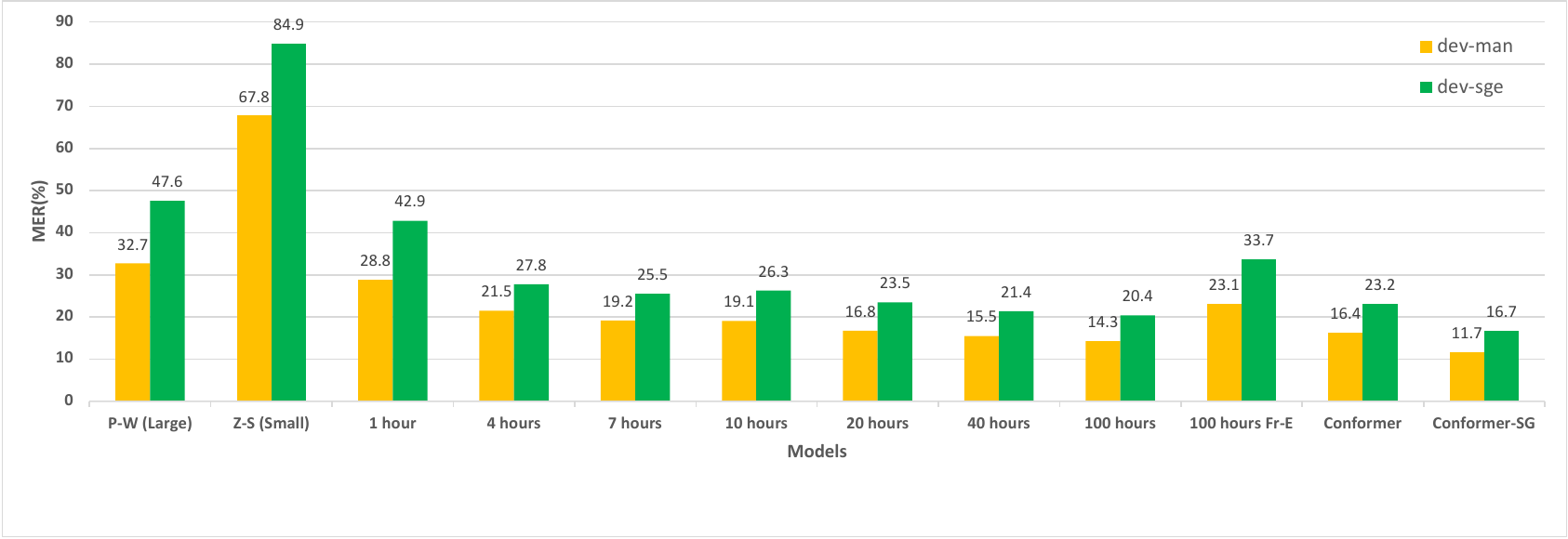}
    \caption{MER(\%) results of SEAME corpus with diverse training data size finetune on Whisper-Small model. \textbf{P-W (Large)} stands for PromptingWhisper Large model, and \textbf{Z-S (Small)} corresponds to directly decode Whisper-Small model with \textbf{Auto} Language Prompt mentioned in \textnormal{Table~\ref{tab:zero-shot-results}}. 
    \{1,4,7,10,20,40,100\} hours indicate the Whisper-Small models that are finetuned with corresponding training data size and \textnormal{\textlangle\textbar en\textbar\textrangle}~is fixed during both finetuning and decoding stages. \textbf{Fr-E} denotes Freeze-Encoder when finetune the Whisper-Small model.  \textbf{Conformer-SG} is our in-house English-Mandarin-Malay trilingual model that was trained with over 40k hours Singaporean Accented data.
    }
    \label{fig:Ab.SEAME}
    \vspace{-0.9em}
\end{figure*}

\begin{figure*}[htb]
    \centering
    \includegraphics[width=0.95\linewidth]{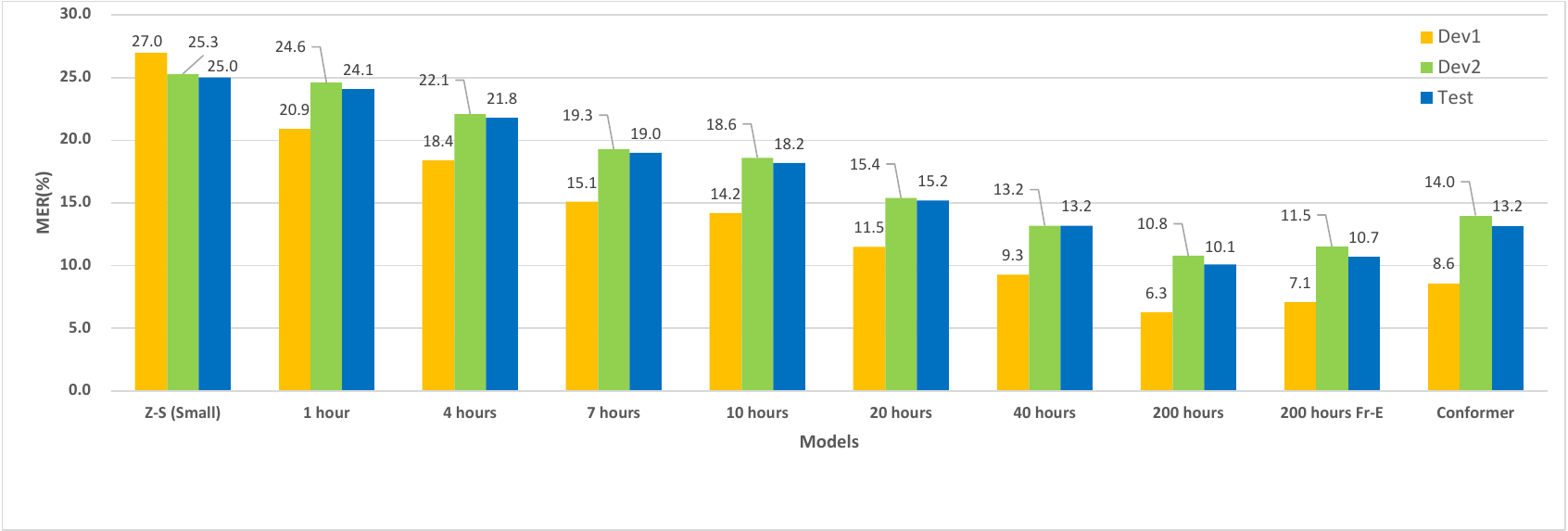}
    \caption{MER(\%) results of ASRU corpus with diverse training data size finetune on Whisper-Small model. \textbf{Z-S (Small)} corresponds to directly decode Whisper-Small model with \textnormal{\textlangle \textbar zh\textbar \textrangle}~Language Prompt. 
    \{1,4,7,10,20,40,200\} hours indicate the Whisper-Small models that are finetuned with corresponding training data size and \textnormal{\textlangle\textbar zh\textbar\textrangle}~is fixed during both finetuning and decoding stages. \textbf{Fr-E} denotes Freeze-Encoder when finetune the Whisper-Small model.
    }
    \label{fig:Ab.ASRU}
    \vspace{-0.9em}
\end{figure*}
    Specifically, \textbf{\textlangle \textbar en\textbar \textrangle} ~prompt achieves best performance for both test sets of SEAME corpus and also introduces $2.3\sim2.9\%$ absolute MER reduction compared with Conformer model, while \textbf{\textlangle \textbar en\textbar \textrangle \textlangle \textbar zh\textbar \textrangle} ~prompt yields optimal results on all Dev and Test sets of ASRU corpus which demonstrates $2.3\sim3.4\%$ absolute MER reduction compared with Conformer model. However, we realize that language prompts, which are either from one of \textbf{\textlangle \textbar en\textbar \textrangle}~and \textbf{\textlangle \textbar zh\textbar \textrangle}~or both, or even our proposed fusion prompt, produce similar results (less than 1\% MER gap among all models for each testset). So we introduce Russian Language Prompt \textbf{\textlangle \textbar ru\textbar \textrangle}~as reference in this experiment. Our findings are confirmed by the results of \textbf{\textlangle\textbar ru\textbar\textrangle} ~which demonstrate that regardless of what language prompt given to the Whisper decoder, once the finetuning process is complete, the performance achieved will be similar.

\section{Ablation Study}
\label{sec:Abl}
In this section, we primarily investigate how the size of the training dataset affects the performance outcomes of fine-tuning the Whisper model. Additionally, we explore the effects on Whisper's performance when the encoder parameters are frozen during the finetuning process and solely update the decoder parameters.

First, for all the finetuned Whisper 
models, we fix the language prompts \textbf{\textlangle \textbar en\textbar\textrangle}~for SEAME and \textbf{\textlangle \textbar zh\textbar\textrangle}~for ASRU2019 experiments. We then randomly subset $(1,4,7,10,20,40)$ hours from both SEAME and ASRU training set, and update the Whisper-Small model for 8000 steps to examine the smallest size of data that could obtain a practical CS-ASR system.
Also, we try to freeze the entire encoder of the Whisper model while performing finetuning on the whole dataset to determine if finetuning the decoder alone can effectively bridge the gap between multilingual and Code-Switching scenarios. Additionally, this can help to verify whether the Whisper Encoder can overcome the acoustic mismatch that often arises in such diverse linguistic environments. 

Figure~\ref{fig:Ab.SEAME} shows the results on SEAME corpus. The results show that only 1 hour training data can produce model outperforms PromptingWhisper-Large by around 10\%, and obtains around 50\% MER reduction. When the amount of training data reaches 20 hours and 40 hours, the model will yield results comparable to or surpass the conformer model that is trained with the entire training set. However, when the encoder parameters are frozen, the performance drop could be at most 65\% which suggest that the encoder of the Whisper-Small model may exhibit a substantial mismatch with SEAME corpus (Singaporean Accents), potentially leading to degradation in its performance. 
In Figure~\ref{fig:Ab.SEAME}, we also present the Conformer-SG, which is an U2++ Conformer streaming ASR model as referenced in~\cite{}. This model, trained on over 40,000 hours of Mandarin-English-Malay speech data, achieves state-of-the-art performance on the SEAME corpus and surpasses the finetuned Whisper-Small model by approximately 18\%.

Figure~\ref{fig:Ab.ASRU} shows the results on ASRU corpus. The results show similar conclusion we obtain from experiments of SEAME corpus. However, the frozen encoder experiment for ASRU corpus shows much tiny performance degradation compared with SEAME corpus, which suggest the performance gap tends to diminish when dealing with scenarios, such as speech without obvious accents, that closely align with the conditions and characteristics of the Whisper model’s training set.
\section{Conclusion}
\label{sec:conclusion}
In this paper, our findings reveal that adapting the Whisper model with code-switching datasets significantly enhances its capability for code-switching speech recognition (CS-ASR), even in contexts with limited resources. Our experiments, which span a variety of linguistic backgrounds, demonstrate that while different prompting strategies yield varied performances prior to adaptation, after adapting the Whisper model with code-switching speech data, these strategies result in similarly enhanced performance, effectively mitigating the complexities inherent in code-switching environments. These adaptations not only bolster the model's overall performance but also align with the broader goal of developing ASR systems that are more inclusive and precise for multilingual users. 
This research thus marks a crucial step forward in understanding the potential of large foundational models for navigating the intricate dynamics of code-switching in various linguistic scenarios.

\bibliographystyle{IEEEtran}
\bibliography{mybib}

\end{document}